\documentclass{article}[11pt]
\usepackage{graphicx}

\begin{document}

\title{\bf A novel approach to background reduction in double beta decay
  experiments} 
\author{T. Tabarelli de Fatis \\
Universit\`a and INFN Milano Bicocca}

\maketitle

{ {\bf ABSTRACT:} Active background reduction in high resolution
  calorimeters is a promising approach to achieve ultimate sensitivity
  in neutrinoless double beta decay experiments. We propose Cerenkov
  emission from beta rays in bolometric crystals as a viable
  alternative to scintillation. This novel approach could broaden the
  range of materials of interest for calorimetric searches of the
  double beta decay. We discuss the optical properties of TeO$_2$
  crystals, as a show case.  }

\section{Introduction}

Neutrinoless double beta decay is a sensitive probe of the charge
conjugation properties of neutrinos and of the neutrino mass
scale. This decay is expected to occur for several even-even nuclei at
a (tiny) rate proportional to the square of the effective Majorana
mass of the electron neutrino. Neutrino oscillations results suggest
that the effective electron neutrino mass be of order 100~meV or
10~meV for inverted or direct hierarchy respectively, unless the 
CP-phases of Majorana neutrinos conspire for cancellations. For this
low mass scales to be tested, sensitivities to half-life time of the 
candidate nuclei of order $10^{25}$~y need to be achieved. 

The experimental signature for the decay is rather clean in itself and 
corresponds to a peak, only smeared by the instrumental resolution,
in the spectrum of the energy sum of the two electrons emitted in the
decay. Yet, the search for such a rare decay is thwarted by the
background of spurious events due to radioactive contaminants in 
the environment and in the detector material. 

The struggle for background reductions in double beta decay experiments
is documented by decades of R\&D projects in several directions,
exploiting a variety of experimental techniques. In addition to
careful selections of materials with low radioactive contaminants,
merit factors in modern experiments are the detector resolution and
dual readout techniques to provide further rejection power. A detailed
comparison of the different techniques is beyond the scope of this
paper and the reader is referred to review articles on the subject
\cite{rassegna}. Here we propose that a dual readout approach based on
the detection of Cerenkov photons emitted by beta rays could win a
sizable background rejection. This option, alternative to
scintillation, may open the window to a new set of interesting
materials for double beta decay searches. As a show case, we discuss a
potential application to the CUORE experiment \cite{CUORE}, a high
resolution bolometric calorimeters based on TeO$_2$ crystals. 

\section{Dual readout approaches in bolometric experiments} 

Bolometers are crystals operated at cryogenic temperatures, where the
low thermal capacitance implies a sizable (measurable) variation of the
crystal temperature upon the energy deposit of a particle. This
technique has long been pioneered with TeO$_2$ crystals and has now
grown into one of the leading experiments in the field. Yet, the 
experiment is plagued by an important background due to alpha decays
with energy deposits non fully contained in a crystal, due to residual
contaminants in the crystal itself, on its surfaces or in the
materials of the mechanical supports. Alpha 
particles depositing only a fraction of their energy in the bolometer
volume give rise to a continuum in the energy spectrum, which reduces
the sensitivity to the the feeble double beta decay signal, expected
at around 2.53~MeV for $^{130}$Te. Detailed analysis of the spurious
events recorded by the CUORE ancestor, the Cuoricino setup
\cite{Cuoricino}, demonstrated that about 2/3 of the background at the
energy of the expected neutrinoless double beta decay transition can
be ascribed to this background.

Several approaches for active background reduction have been
proposed, based on the anti-coincidence of nearby crystals or on the
use of designated veto Si bolometers mounted on the TeO$_2$ surfaces
\cite{Pedretti}.  
Alternatively, a dual readout approach based on the simultaneous
measurement of the bolometric signals and of the scintillation light
has been advocated, with applications both in double beta decay and in
Dark Matter searches \cite{Aless,CRESST}. The rational of this
technique is that the scintillation light due to heavy particles is
strongly quenched and the measurement of the ratio of the emitted
light over the deposited energy allows for signal/background
discrimination. While recent results in this direction are encouraging   
on some crystals \cite{Pirro}, this approach is only limited to
materials that exhibit some scintillation property. In particular, it
was not attempted with TeO$_2$, which is not classified as a
scintillator. Still, TeO$_2$ crystals do have suitable optical
properties to act as excellent Cerenkov radiators, as we specify in
the next section. Similarly to scintillation, we argue that Cerenkov
light emitted by fast electrons can be exploited to tag double beta 
decay events against the alpha background. 

\section{A show case: the Cerenkov yield from neutrino-less double
  beta decay of Te-130} 

TeO$_2$ is a clear crystal, transparent to light from about 350~nm to
the infrared region, and it has index of refraction $n=2.4$,
corresponding to a threshold for Cerenkov emission of about 50~keV for
electrons and about 400~MeV for alpha particles. These yields imply
that the use of Cerenkov radiation to tag double beta decays offers
100\% rejection power against alpha emitters contaminating the
detector. Random coincidences in a low background experiment are
negligible.  

\begin{figure}[htb]
\begin{center}
\includegraphics[width=0.95\textwidth]{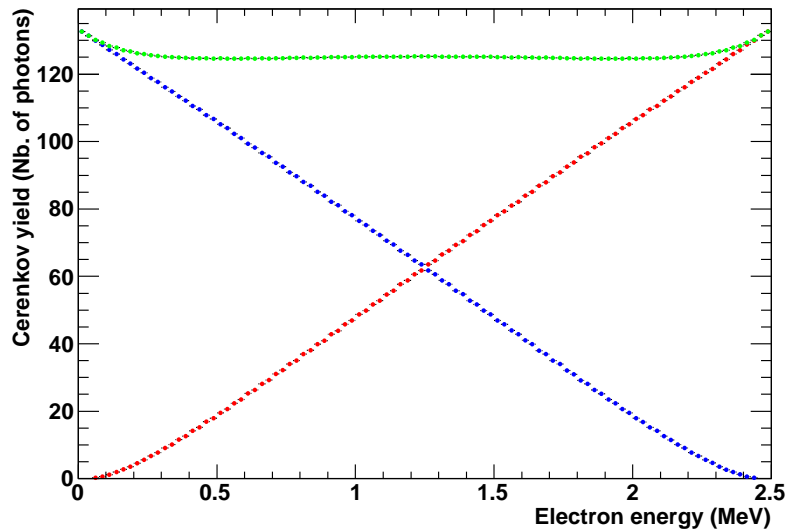}
\caption{Calculated Cerenkov yield from 350~nm to 600~nm in TeO$_2$ as
  a function of the kinetic energy of one of the two electrons emitted
  in the $\beta\beta$-decay of $^{130}$Te. The three curves show the
  yield associated to that electron (red), the yield of the other
  electron (blue) and the total yield from the sum of the two
  electrons. } 
\label{CerYield}
\end{center}
\end{figure}

A precise estimate of the Cerenkov yield from the two electrons
emitted in the neutrinoless double beta decay of $^{130}$Te was
derived from the above-mentioned optical properties of the crystal and
from electron range tables for a 6~g/cm$^3$ density TeO$_2$ crystal
\cite{NIST}. The energy spectrum of the two electrons in the pair has
been simulated according to the parametrisation $dN/dT_e \propto 
(T_e+1)^2(Q+1-T_e)^2$, where $T_e$ is the kinetic energy of the
electron and $Q$ is the transition energy \cite{Pr-Rosen}. From the
above, we calculate that the two electrons emit in total along their
range about 125 Cerenkov photons in an energy interval from 2 eV
(about 600 nm) to 3.5 eV (about 350 nm). Fluctuations around this
value are expected from Poisson statistics and from the sharing of the
energy between the two electrons. As we show in Fig.\ref{CerYield},
the Cerenkov yields from the sum of the two electrons is practically
independent of the energy sharing between the two electrons. The
relative yield increases by less than about 10\%, when one of the
electrons takes the entire decay energy. This figure is comparable or
lower than the expected stochastic fluctuations associated to light
emission and detection.  

On the basis of the curves shown in Fig.\ref{CerYield}, one might
argue that Cerenkov emission could also provide some discrimination
power against gamma-induced background events. In this case, at
variance with beta rays, the energy is deposited through multiple
interactions, in which the photon energy is transferred to low energy
electrons. For multiplicity $N$, a Cerenkov yield equivalent to the
one from a single electron of kinetic energy equal to 
$E_{\gamma}-(N-1) \times E_{th}$, where $E_{\gamma}$ and $E_{th}$ are the
photon energy and the effective energy threshold for Cerenkov
emission, is expected. We have not addressed this in a detailed study,
but simple arguments based on Fig.\ref{CerYield} indicate that any
potential discrimination power is spoiled by stochastic fluctuations,
when light collection and quantum efficiencies are considered.  

Light collection and quantum efficiency may constitute a serious
experimental challenge at cryogenic temperatures. Yet, the conspicuous
Cerenkov yield could offer full rejection of the alpha background even
in the limit of very low efficiencies: a positive tag of a double beta
decay would indeed be provided even by the detection of one single
photoelectron! 

Experimental aspects on the light detection at low temperatures are
discussed in the literature ~\cite{Aless,CRESST,Pirro}. While
conventional Si photon-detectors can easily reach single photoelectron
sensitivities at room temperature, early attempts to operate these
devices at cryogenic temperatures were abandoned, because of technical
difficulties \cite{Aless}. Alternatively, light absorbers operated as
bolometers can be used. The optimization of such devices for Dark
Matter searches has already reached sensitivities to light yields
comparable or lower to the ones predicted in this paper for the
Cerenkov emission associated to double beta decays \cite{CRESST}. In
conclusion, although a dedicated R\&D work may be needed for the
specific case discussed in this paper, existing experimental
techniques seem to offer already the required sensitivity.   

\section{Summary}

We propose that the use of Cerenkov radiation emitted by the two
electrons from neutrinoless beta decay can be used for active
background reduction purposes in double beta decay experiments. 
This technique is alternative to active background reduction
approaches based on the scintillation properties of materials, and
could broaden the set of interesting materials for double beta decay
experiments. Particularly interesting would be the case of nuclei
with double beta transition energy above 2.614~MeV: the energy of the
highest gamma ray from natural radioactivity. As a show case, we have
discussed instead the properties of TeO$_2$ crystals, adopted by the
CUORE experiment, plagued by alpha induced spurious events. We
estimate a total yield of about 125 Cerenkov photons per neutrinoless
double beta decay, prior to quantum efficiency and light collection
effects. This light emission is detectable by available experimental
techniques. In particular, low light level detectors developed for
Dark Matter searches seem to match the required performance. 

\section*{Acknowledgements}
Discussions with S.Ragazzi and O.Cremonesi are acknowledged.


\begin{thebibliography}{9}

\bibitem{rassegna} See for example F.T.Avignone {\it et al.}, Rev. Mod. Phys. 80, 481 (2008). 
\bibitem{CUORE} R.Ardito {\it et al.}, The CUORE Collaboration, Prog. in Part. and Nucl. Phys. 57 (2006) 2003. 
\bibitem{Cuoricino} C.Arnaboldi {\it et al.}, Phys. Rev. C78, 035502, (2008). 
\bibitem{Pedretti} M.Pedretti {\it et al.}, AIP Conf. Proc. 897: 59-64, 2007 
\bibitem{Aless} A.Alessandrello {\it et al.}, Nucl. Phys. B (proc. Suppl.) 28A (1992), 233; 
                A.Alessandrello {\it et al.}, IEEE Trans Nucl. Sci. 39-4 (1992) 612.
\bibitem{CRESST} C. Bobin {\it et al}, Nucl. Instr. and Meth. A386 (1994) 453;
                 F.Petricca {\it et al.}, Nucl. Instr. and Meth. A520 (2004) 193; 
                 G.Angloher {\it et al.}, Astropart. Phys. 23 (2005) 325; 
\bibitem{Pirro} S.Pirro {\it et al.}, Phys. Atomic Nucl. 69 (2006) 2109; 
                L. Gironi {\it et al.}, arXiv:nucl-ex/08095126
\bibitem{NIST} ``Stopping-power and range tables for electrons'', ICRU Report No. 37 (1984). 
\bibitem{Pr-Rosen} H.Primakoff and S.Rosen, Rep. Prog. Phys. 22 (1959) 121-166. 

\end{thebibliography}
\end{document}